\documentclass[12pt,reqno]{amsart}
\usepackage{graphics}
\usepackage{subfigure}
\usepackage{psfrag}
\usepackage{epsfig}
\usepackage{pstricks}
\usepackage{amsthm}
\usepackage{amsmath}
\usepackage{amsfonts}
\usepackage{ae}
\usepackage{graphicx}

\newtheorem{theorem}{Theorem}[section]

\theoremstyle{definition}

\theoremstyle{remark}
\newtheorem{remark}[theorem]{Remark}

\numberwithin{equation}{section}

\begin{document}

\title[Nonlinear Neutral Inclusions]{Nonlinear Neutral Inclusions:\\ Assemblages of Spheres}

\author[Silvia Jim\'enez \and Bogdan Vernescu \and William Sanguinet]{Silvia Jim\'enez \and Bogdan Vernescu \and William Sanguinet\\\\\tiny{Dept. of Mathematical Sciences, Worcester Polytechnic Institute\\100 Institute Road, Worcester, MA 01609-2280 USA\\Phone: +1-508-831-5241 | Fax: +1-508-831-5824}\\\tiny{E-mail: silviajimenez@wpi.edu (Jim\'enez), vernescu@wpi.edu (Vernescu)}, wcsanguinet@wpi.edu (Sanguinet)}
\textit{}
\begin{abstract}
If a neutral inclusion is inserted in a matrix containing a uniform applied electric field, it does not disturb the field outside the inclusion. The well known Hashin coated sphere is an example of a neutral coated inclusion. In this paper, we consider the problem of 
constructing neutral inclusions from nonlinear materials. In particular, we discuss assemblages of coated spheres and the two-dimensional analogous problem of assemblages of coated disks.  
\end{abstract}
\maketitle
\section{Introduction}
A neutral inclusion, when inserted in a matrix containing a uniform applied electric field, does not disturb the field outside the inclusion.  The problem of finding neutral inclusions goes back to 1953 when Mansfield found that certain reinforced holes, which he called ``neutral holes", could be cut out of a uniformly stressed plate without disturbing the surrounding stress field in the plate \cite{Mansfield1953}.  The analogous problem of a ``neutral elastic inhomogeneity" in which the introduction of the inhomogeneity into an elastic body (of a different material), does not disturb the original stress field in the uncut body, was first studied by Ru \cite{Ru1998}.
	
The well known Hashin coated sphere is an example of a neutral coated inclusion \cite{Hashin1962}.   In \cite{Hashin1962b,Hashin1963}, Hashin and Shtrikman found an exact expression for the effective conductivity of the coated sphere assemblage.  For the case of imperfect interfaces, neutral spherical inclusions have been studied in \cite{Lipton1995}, \cite{Lipton1996}, \cite{Lipton1996-2}, \cite{Lipton1997}, \cite{Lipton1997-2}, and \cite{Lipton1999}.  Examples of neutral inclusions of arbitrary shape have been studied in \cite{Milton2001}.  For other references see \cite{Vernescu2010} and Chapter~$7$ of \cite{Milton2002}.  

\begin{figure}[h]
\centering
\includegraphics[width=0.3\textwidth]{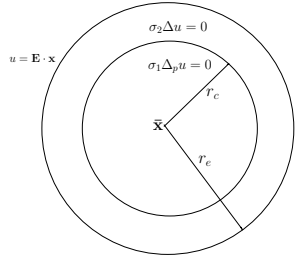}
\caption{Cross section of a coated sphere}
\label{fig:disk}
\end{figure}

To begin, we consider a particular coated sphere (see Fig.~(\ref{fig:disk})), of phase $1$ in the core and phase $2$ in the coating, of core radius $r_c$ and exterior radius $r_e$, $1<r_{c}<r_{e}$, subject to linear boundary conditions, i.e. we apply the linear field $\mathbf{E}\cdot\mathbf{x}=Ex_{1}$, (where $\mathbf{E}=E\mathbf{e^{1}}$, with $\mathbf{e^{1}}=(1,0,0)$ and $\mathbf{x}=(x_{1},x_{2},x_{3})$) as a boundary condition to the exterior boundary of the sphere, to find a solution $u$ that solves
	\begin{equation}
		\label{PDENeutralInclusions}
		\begin{cases}
			\sigma_{1}\Delta_{p}u=\sigma_{1}\nabla\cdot\left(\left|\nabla u\right|^{p-2}\nabla u\right)=0 \text{ (nonlinear) in the core,}\\
			\sigma_{2}\Delta u=\sigma_{2}\nabla\cdot\nabla u=0 \text{ (linear) in the coating}
		\end{cases}
	\end{equation}
where the conductivities $\sigma_{i}$ for $i=1,2$ satisfy $\infty>\sigma_{1}>\sigma_{2}>0$ and $\Delta_{p}$ represents the $p$-Laplacian ($p>1$), together with the usual compatibility conditions at the interfaces (continuity of the electric potential and continuity of the normal component of the current).  The ratio of the inner sphere radius to the outer sphere radius is fixed, therefore the volume fractions of both materials $\theta_{1}$ and $\theta_{2}$ are fixed:
	\begin{equation}
		\label{volumefraction}
		\theta_{1}=\frac{4/3\pi r_c^3}{4/3\pi r_{e}^3}=\frac{r_c^3}{r_{e}^3}\text{, \hspace{3mm}and \hspace{3mm}$\theta_{2}=1-\theta_{1}$.}
	\end{equation}

\begin{figure}[h]
\centering
\includegraphics[width=0.3\textwidth]{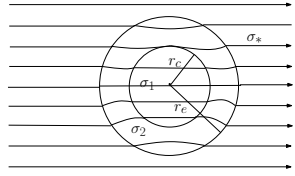}
\caption{When $\sigma_*$ is chosen appropriately, one can insert a coated sphere into the medium without disturbing the surrounding uniform current field}
\label{fig:coated}
\end{figure}

\begin{figure}[h]
\centering
\includegraphics[width=0.3\textwidth]{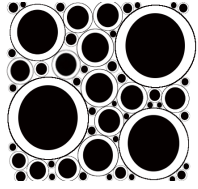}
\caption{Cross section of the assemblage of coated spheres.}
\label{fig:assemblage}
\end{figure}

Our calculations in Section~\ref{Stat-Spheres} show that we can replace the coated sphere with a sphere composed only of linear material of conductivity $\sigma_{*}$ (see (\ref{7}) and Fig.~\ref{fig:coated}).  Since the equations for conductivity are local equations, one could continue to add similar coated sphere of various sizes without disturbing the prescribed uniform applied field surrounding the inclusions.  In fact, we can fill the entire space (aside from a set of measure zero) with a periodic assemblage of these coated spheres by adding coated spheres of various sizes ranging to the infinitesimal and it is assumed that they do not overlap the boundary of the unit cell of periodicity (see Fig~(\ref{fig:assemblage})).  The spheres can be of any size, but the volume fraction of the core and the coating layer is the same for all spheres.  While adding the coated spheres, the flux of current and electrical potential at the boundary of the unit cell remains unaltered.  Therefore, the effective conductivity does not change.  

This configuration of nonlinear materials dissipates energy the same as a linear material with thermal conductivity $\sigma_{*}$ (see (\ref{7})).
  
This paper is structured as follows: Section~\ref{Stat-Spheres} provides the statement of the problem and the main result for an assemblage of coated spheres, Section~\ref{Stat-Disks} extends the results for an assemblage of coated disks, Section~\ref{sigma} contains a discussions about the effective conductivity of the assemblage and its relation to $p$, and Section~\ref{Fin} contains the conclusions and a description of future work.

\section{Assemblage of Coated Spheres: Statement of the Problem and Result}  
\label{Stat-Spheres}

Let $\mathbf{\bar{x}}$ be the center of the coated sphere (See Fig.~(\ref{fig:disk})).  Inside the sphere, we ask that 
\begin{equation}
	\label{PDENeutralInclusions2}
	\begin{cases}
		\sigma_{1}\Delta_{p}u=0 & 0<\left|\mathbf{x}-\mathbf{\bar{x}}\right|<r_{c} \\
		\sigma_{2}\Delta u=0 & r_c<\left|\mathbf{x}-\mathbf{\bar{x}}\right|<r_e, \\
	\end{cases}
\end{equation}
where $\sigma_{1}$ and $\sigma_{2}$ are positive constants, together with the usual compatibility conditions at the interfaces: 

\begin{equation}
	\label{conta}
	\text{$u$ continuous across $\left|\mathbf{x}-\mathbf{\bar{x}}\right|=r_c$,}
\end{equation}
\begin{equation}
	\label{extbound}
	\text{$u=Ex_{1}$ at $\left|\mathbf{x}-\mathbf{\bar{x}}\right|=r_e$,}
\end{equation}
and the transmission conditions
\begin{equation}
	\label{transa}
	\text{$\sigma_{1}\mathbf{n}\cdot\left|\nabla u\right|^{p-2}\nabla u=\sigma_{2}\mathbf{n}\cdot\nabla u$,  across $\left|\mathbf{x}-\mathbf{\bar{x}}\right|=r_c$,}
\end{equation}
and
\begin{equation}
	\label{transb}
	\text{$\sigma_{2}\mathbf{n}\cdot\nabla u=\sigma_{*}\mathbf{n}\cdot\nabla u$,  across $\left|\mathbf{x}-\mathbf{\bar{x}}\right|=r_e$.}
\end{equation}

Using polar coordinates, we look for a solution $u$ of (\ref{PDENeutralInclusions2}) satisfying (\ref{conta})-(\ref{transb}), with radial symmetry, of the form 
\begin{equation}
	\label{u-ni}
	\begin{cases}
	\displaystyle
		u=a_{1}r\cos\theta & \text{ for $0<r<r_c$,}\\
		u=\frac{b_{2}}{r^2}\cos\theta+a_{2}r\cos\theta & \text{ for $r_c<r<r_e$,}\\
		u=Er\cos\theta & \text{ for $r=r_e$.}
	\end{cases}
\end{equation}
where $r=\left|\mathbf{x}-\mathbf{\bar{x}}\right|$ and $\theta$ measures the angle between the unit vector $\mathbf{v}$ in the direction of the applied field and $\mathbf{x}$.

We introduce the notation for the unit radial vector $\displaystyle \mathbf{e_{r}}=\mathbf{n}=\frac{\mathbf{x}-\mathbf{\bar{x}}}{\left|\mathbf{x}-\mathbf{\bar{x}}\right|},$ and denote by $\mathbf{e_{\theta}}$ the unit vector perpendicular to $\mathbf{e_{r}}$.  We have $\mathbf{v}=\cos\theta\mathbf{e_{r}}-\sin\theta\mathbf{e_{\theta}}$.  

Since (\ref{u-ni}) satisfies (\ref{PDENeutralInclusions2}), it is left to show that it satisfies the required compatibility conditions at the interfaces, i.e., we require that the solution $u$ of (\ref{PDENeutralInclusions2}) satisfies (\ref{conta})-(\ref{transb}).

In what follows, we explain how the unknowns $a_{1}$, $a_{2}$, and $b_{2}$ and the effective conductivity $\sigma_{*}$ are determined from (\ref{conta}), (\ref{extbound}), (\ref{transa}), and (\ref{transb}).

First, we look at the conditions $u$ must satisfy when $r=r_c$.  From (\ref{conta}), we have
\begin{equation}
	\label{1}
	\displaystyle
	a_{1}=a_{2}+\frac{b_{2}}{r_{c}^{3}}, 
\end{equation}
and from (\ref{transa}), we obtain
\begin{equation}
	\label{2}
	\displaystyle
\sigma_{1}\left|a_{1}\right|^{p-2}a_{1}=\sigma_{2}\left(a_{2}-\frac{2b_{2}}{r_c^3}\right). 
\end{equation}

We now look at the conditions that $u$ must satisfy when $r=r_e$.  From (\ref{extbound}), we have
\begin{equation}
	\label{3}
	\displaystyle
	E=a_{2}+\frac{b_{2}}{r_{e}^3}. 
\end{equation}

and from (\ref{transb}), we obtain
\begin{equation}
	\label{4}
	\displaystyle
	\sigma_{2}\left(a_{2}-\frac{2b_{2}}{r_{e}^3}\right)=\sigma_{*}E. 
\end{equation}

Eliminating $a_{2}$ from (\ref{1}) and (\ref{3}), we obtain
\begin{equation}
	\label{5}
	\displaystyle a_{1}=E+\frac{b_{2}}{r_e^3}\left(\frac{r_e^3}{r_{c}^3}-1\right).
\end{equation}

Let us denote $$A=\frac{r_e^3}{r_{c}^3}-1=\frac{1}{\theta_1}-1$$ and $$ B=\frac{2r_e^3}{r_{c}^3}+1=\frac{2}{\theta_1}+1.$$

Observe that both $A>0$ and $B>0$ are independent of $r_c$ and $r_e$, i.e. they are defined only in terms of $\theta_1$. 

Using  (\ref{2}) and (\ref{3}) in (\ref{5}), we obtain the following identity
\begin{align}
	\label{6}
&\sigma_{1}\left|E+\frac{b_{2}}{r_e^3}\left(\frac{r_e^3}{r_{c}^3}-1\right)\right|^{p-2}\left[E+\frac{b_{2}}{r_e^3}\left(\frac{r_e^3}{r_{c}^3}-1\right)\right]=\sigma_{2}\left(E-\frac{b_{2}}{r_{e}^3}-\frac{2b_{2}}{r_c^3}\right), \notag
	&\intertext{which can be rewritten in terms of $A$ and $B$ as}
	&  \sigma_{1}\left|E+A\left(\frac{b_{2}}{r_e^3}\right)\right|^{p-2}\left[E+A\left(\frac{b_{2}}{r_e^3}\right)\right]-\sigma_{2}\left(E-B\left(\frac{b_{2}}{r_e^3}\right)\right)=0
\end{align}

At this point, we consider the function 
\begin{equation}
\label{f}
\displaystyle
f(x)=\sigma_{1}\left|E+Ax\right|^{p-2}(E+Ax)-\sigma_{2}(E-Bx).
\end{equation}
Note that we obtain $b_{2}$ if we can prove that $f(x)=0$ has a (unique) solution.  If that is the case, from (\ref{3}) and (\ref{5}), we can obtain $a_{1}$ and $a_{2}$ and from  (\ref{4}), we can get an expression for $\sigma_{*}$.

Let us study $f(x)$.  If $E+Ax\geq0$, we have $$f(x)=\sigma_{1}(E+Ax)^{p-1}-\sigma_{2}(E-Bx).$$  Taking the derivative of the $f(x)$, we have $$f'(x)=A\sigma_{1}(p-1)(E+Ax)^{p-2}+\sigma_{2}B$$ which is positive for all $x$, so the function $f(x)$ is increasing.

If $E+Ax<0$, we have $$f(x)=-\sigma_{1}(-E-Ax)^{p-1}-\sigma_{2}(E-Bx),$$ and here $$f'(x)=A\sigma_{1}(p-1)(-E-Ax)^{p-2}+\sigma_{2}B$$ is positive for all $x$ so the function $f(x)$ is also increasing in this case.

Observe that as $x\rightarrow-\infty$, the function $f(x)$ approaches $-\infty$ and as $x\rightarrow\infty$, the function $f(x)$ approaches $\infty$.  

Therefore, we conclude that $f(x)$ has a unique solution $x_{0}$.  Moreover, observe that the coefficients of $f(x)$ depend only on $\sigma_{1}$, E, $\sigma_{2}$, $\theta_{1}$, and $p$, then  
\begin{equation}
	\label{ind}
	\displaystyle x_{0}=\frac{b_{2}}{r_e^3} =K(\sigma_{1},\sigma_{2},E,\theta_{1},p).
\end{equation}

Consequently, from (\ref{ind}) we have $b_{2}=x_{0}r_e^3$ which allows us to obtain $a_{2}$ and $a_{1}$ from  (\ref{3}) and (\ref{5}), respectively.  To obtain an expression for $\sigma_{*}$ we use  (\ref{3}) in (\ref{4}) as follows
\begin{align}
	\label{7}
	\displaystyle
	&  \sigma_{*}=\frac{\sigma_{2}}{E}\left(E-3\frac{b_{2}}{r_{e}^3}\right),\notag
	&\intertext{ and from (\ref{ind}), we get}
	& \sigma_{*}=\frac{\sigma_{2}}{E}\left(E-3K(\sigma_{1},\sigma_{2},E,\theta_{1},p)\right)		 
\end{align}

Here, we would like to emphasize that (\ref{7}) shows that $\sigma_{*}$ does not depend on $r_c$ or $r_e$.  Therefore, we realize that $\sigma_{*}$ is independent of scale.  The rest of the coated spheres inserted into the matrix are chosen identical up to a scale factor, so that when inserted, the field is not disturbed.  In this way, $\theta_{1}$ is the proportion of nonlinear material in the resulting assemblage of coated spheres and $\sigma_{*}$ is its effective conductivity.

\begin{remark}
If $p=2$, $$f(x)=\sigma_{1}(E+Ax)-\sigma_{2}(E-Bx)$$ has a unique root $\displaystyle x_{0}=\frac{(\sigma_{2}-\sigma_{1})E}{A\sigma_{1}+B\sigma_{2}}$ and in this case 
\begin{align}
	\label{HSformula}
	\displaystyle
\sigma_{*}&=\frac{\sigma_{2}}{E}\left(E-\frac{3(\sigma_{2}-\sigma_{1})E}{A\sigma_{1}+B\sigma_{2}}\right)\notag\\
&=\sigma_{2}+\frac{3\theta_{1}\sigma_{2}(\sigma_{1}-\sigma_{2})}{3\sigma_{2}+\theta_{2}(\sigma_{1}-\sigma_{2})}
\end{align}  
which is the Hashin-Shtrikman formula.  
\end{remark}

\section{Assemblage of Coated Disks: Statement of the Problem and Result}  
\label{Stat-Disks}

Following the same method, we obtained the same results in the two-dimensional case of assemblages of coated disks.  Consider a particular coated disk of center $\mathbf{\bar{x}}$ with core radius $r_c$ and exterior radius $r_e$, $1<r_{c}<r_{e}$, subject to linear boundary conditions $\mathbf{\tilde E}\cdot\mathbf{x}=\tilde{E}x_{1}$, (where $\mathbf{\tilde E}=\tilde E\mathbf{e^{1}}$, with $\mathbf{e^{1}}=(1,0)$ and $\mathbf{x}=(x_{1},x_{2})$) as a boundary condition to the exterior boundary of the disk, to find a solution $u$ that solves (\ref{PDENeutralInclusions}) together with the usual compatibility conditions at the interfaces.  The ratio of the inner disk radius to the outer disk radius is fixed, therefore the area fractions of both materials $\tilde\theta_{1}$ and $\tilde\theta_{2}$ are fixed:
	\begin{equation}
		\label{areafraction}		\tilde\theta_{1}=\frac{r_c^2\pi}{r_{e}^2\pi}=\frac{r_c^2}{r_{e}^2}\text{, \hspace{3mm}and \hspace{3mm}$\tilde\theta_{2}=1-\tilde\theta_{1}$.}
	\end{equation}

Accordingly, using polar coordinates, we look for a solution $u$ of (\ref{PDENeutralInclusions2}) satisfying (\ref{conta})-(\ref{transb}), with radial symmetry, of the form 
\begin{equation}
	\label{u-nid}
	\begin{cases}
	\displaystyle
		u=\tilde a_{1}r\cos\theta & \text{ for $0<r<r_c$,}\\
		u=\frac{\tilde b_{2}}{r}\cos\theta+\tilde a_{2}r\cos\theta & \text{ for $r_c<r<r_e$,}\\
	\end{cases}
\end{equation}
with $r$, $\theta$, $\mathbf{e_{r}}$ and $\mathbf{e_{\theta}}$ described in a similar way as in the previous section.    

The corresponding equations to (\ref{1})-(\ref{4}) are 
\begin{equation}
	\label{1d}
	\displaystyle
	\tilde a_{1}=\tilde a_{2}+\frac{\tilde b_{2}}{r_{c}^{2}}, 
\end{equation}
\begin{equation}
	\label{2d}
	\displaystyle
\sigma_{1}\left|\tilde a_{1}\right|^{p-2}\tilde a_{1}=\sigma_{2}\left(\tilde a_{2}-\frac{\tilde b_{2}}{r_c^2}\right). 
\end{equation}
\begin{equation}
	\label{3d}
	\displaystyle
	\tilde E=\tilde a_{2}+\frac{\tilde b_{2}}{r_{e}^2}. 
\end{equation}
\begin{equation}
	\label{4d}
	\displaystyle
	\sigma_{2}\left(\tilde a_{2}-\frac{\tilde b_{2}}{r_{e}^2}\right)= \sigma_{*}\tilde E. 
\end{equation}

Eliminating $\tilde a_{2}$ from (\ref{1d}) and (\ref{3d}), we obtain
\begin{equation}
	\label{5d}
	\displaystyle \tilde a_{1}=\tilde E+\frac{\tilde b_{2}}{r_e^2}\left(\frac{r_e^2}{r_{c}^2}-1\right).
\end{equation}

Let us denote $$C=\frac{r_e^2}{r_{c}^2}-1=\frac{1}{\tilde\theta_1}-1 \hspace{5mm}\text{and}\hspace{5mm} D=\frac{r_e^2}{r_{c}^2}+1=\frac{1}{\tilde\theta_1}+1.$$

Observe that $C>0$, $D>0$ and both are independent of $r_c$ and $r_e$ (they are defined only in terms of $\tilde \theta_1$). 

Using  (\ref{2d}) and (\ref{3d}) in (\ref{5d}), we obtain the following identity
\begin{equation}
	\label{6d}
\sigma_{1}\left|\tilde E+C\left(\frac{\tilde b_{2}}{r_e^2}\right)\right|^{p-2}\left[\tilde E+C\left(\frac{\tilde b_{2}}{r_e^2}\right)\right]-\sigma_{2}\left(\tilde E-D\left(\frac{\tilde b_{2}}{r_e^2}\right)\right)=0
\end{equation}

At this point, we consider the function $$g(x)=\sigma_{1}\left|\tilde E+Cx\right|^{p-2}(\tilde E+Cx)-\sigma_{2}(\tilde E-Dx).$$
We conclude that $g(x)$ has a unique solution $\tilde x_{0}$.  Moreover, observe that the coefficients of $g(x)$ depend only on $\sigma_{1}$, $\sigma_{2}$, $\tilde E$, p, and $\tilde\theta_{1}$, then  
\begin{equation}
	\label{indd}
	\displaystyle \tilde x_{0}=\frac{\tilde b_{2}}{r_e^2} =\tilde K(\sigma_{1},\sigma_{2},\tilde E,p,\tilde\theta_{1}).
\end{equation}

Consequently, from (\ref{indd}) we have $\tilde b_{2}=\tilde x_{0}r_e^2$ which allows us to obtain $\tilde a_{2}$ and $\tilde a_{1}$ from  (\ref{3d}) and (\ref{5d}), respectively, and 
\begin{equation}
	\label{7d}
\sigma_{*}=\frac{\sigma_{2}}{\tilde E}\left(\tilde E-2\tilde K(\sigma_{1},\sigma_{2},\tilde E,p,\tilde\theta_{1})\right)		 
\end{equation}

\begin{remark}
If $p=2$, $$g(x)=\sigma_{1}(E+Cx)-\sigma_{2}(E-Dx)$$ has a unique root $\displaystyle \tilde{x}_{0}=\frac{(\sigma_{2}-\sigma_{1})E}{C\sigma_{1}+D\sigma_{2}}$ and in this case 
\begin{align}
	\label{HSformula2}
	\displaystyle
\sigma_{*}&=\frac{\sigma_{2}}{E}\left(E-\frac{2(\sigma_{2}-\sigma_{1})E}{C\sigma_{1}+D\sigma_{2}}\right)\notag\\
&=\sigma_{2}+\frac{2\theta_{1}\sigma_{2}(\sigma_{1}-\sigma_{2})}{2\sigma_{2}+\theta_{2}(\sigma_{1}-\sigma_{2})}
\end{align}  
which is the Hashin-Shtrikman formula.  
\end{remark}
\begin{remark}
	When $p=2$, the effective conductivity $\sigma_{*}$ of any isotropic composite of phases $1$ and $2$ (both linear in this case) satisfies the Hashin-Shtrikman bounds (\cite{Hashin1962b,Milton2002}) $$\sigma_{1}+\frac{d\theta_{2}\sigma_{1}(\sigma_{2}-\sigma_{1})}{d\sigma_{1}+\theta_{1}(\sigma_{2}-\sigma_{1})}\geq\sigma_{*}\geq\sigma_{2}+\frac{d\theta_{1}\sigma_{2}(\sigma_{1}-\sigma_{2})}{d\sigma_{2}+\theta_{2}(\sigma_{1}-\sigma_{2})},$$ where $d=2,3$ is the dimensionality of the composite and it is assumed that the phases have been labeled so that $\sigma_{1}>\sigma_{2}$.  Thus, for $p=2$, the coated sphere (\ref{HSformula}) and coated disk assemblages (\ref{HSformula2}) with phase $1$ as core and phase $2$ as coating attain the lower bound and with the phases inverted, attain the upper bound.  They are examples of isotropic materials that, for fixed volume fractions $\theta_1$ and $\theta_2=1-\theta_1$, attain the minimum or maximum possible effective conductivity.  
\end{remark}

\section{Discussion on $\sigma_{*}$ and $x_{0}$}
\label{sigma}
	In this section, we discuss and analyze $\sigma_*$ and $x_{0}$ and also their behavior with respect to $p$.  
	
	First, observe that if we evaluate the function $f$ (\ref{f}) at  $x=-\frac{E}{A}$, we obtain $$f\left(-\frac{E}{A}\right)=-\sigma_{2}\left(E-B\left(\frac{-E}{A}\right)\right)=-\sigma_{2}\left(E+\frac{EB}{A}\right)<0;$$  and if we evaluate $f$ (\ref{f}) at $x=\frac{E}{B}$, we obtain 
	\begin{align*}
		\displaystyle f\left(\frac{E}{B}\right)&=\sigma_{1}\left|E+A\left(\frac{E}{B}\right)\right|^{p-2}\left(E+A\left(\frac{E}{B}\right)\right)\\
		&=\sigma_{1}\left(E+\left(\frac{EA}{B}\right)\right)^{p-1}>0.
	\end{align*}
	Therefore, we can conclude that $x_{0}$ (\ref{ind}) satisfies \begin{equation}
	\label{boundsx0}
	\displaystyle -\frac{E}{A}<x_{0}<\frac{E}{B}, 
\end{equation}
which implies $E+Ax_{0}>0$ and $E-Bx_0>0$; and the effective conductivity $\sigma_{*}$ (\ref{7}) satisfies
\begin{align*}	&\sigma_{2}\left(1-\frac{3}{B}\right)\leq\sigma_{*}\leq\sigma_{2}\left(1+\frac{3}{A}\right)
\intertext{or equivalently}
&\sigma_{2}-\frac{3\sigma_2\theta_{1}}{3-\theta_2}\leq\sigma_{*}\leq\sigma_{2}+\frac{3\sigma_2\theta_1}{\theta_2}
\end{align*}
where $A=\frac{1}{\theta_1}-1>0$ and $ B=\frac{2}{\theta_1}+1>0$.  If $\theta_{1}=1$ (the sphere is made only of nonlinear material), we have $A=0$, $B=3$, and $f(x)=\sigma_{1}\left|E\right|^{p-2}(E)-\sigma_{2}(E-3x)$; from where $x_{0}=\frac{(\sigma_2-\sigma_1\left|E\right|^{p-2})E}{3\sigma_2}$ and $\sigma_{*}=\sigma_{1}\left|E\right|^{p-2}$.  If $\theta_1=0$, from above we obtain $\sigma_*=\sigma_2$.

Let us study the case when $E=1$.  Observe that $f(0)=\sigma_{1}-\sigma_2>0$, therefore we can improve (\ref{boundsx0}) and we obtain 
\begin{equation}
	\label{boundsx01}
	\displaystyle -\frac{1}{A}<x_{0}<0, 
\end{equation}
which implies $0<1+Ax_{0}<1$.  Note that, in this case, $$\frac{\partial x_{0}}{\partial p}=\frac{-\sigma_{1}(1+Ax_{0})^{p-1}\ln(1+Ax_{0})}{\sigma_1(p-1)A(1+Ax_{0})^{p-2}+\sigma_2B}\geq0,$$ and it is equal to $0$ if $A=0$ which corresponds to the case when $\theta_{1}=1$.  So $x_{0}$ is an strictly increasing function of $p$, except when $\theta_{1}=1$ (all nonlinear material), in which case it is a constant function $x_{0}=\frac{\sigma_2-\sigma_1}{3\sigma_2}$.  

\begin{figure}[h]
\centering
\includegraphics[width=0.5\columnwidth]{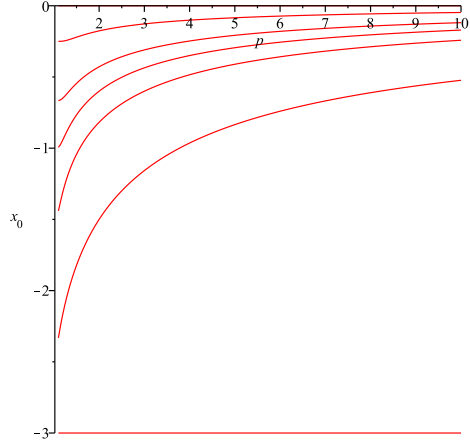}
\caption{Plot of $x_{0}$ with respect to $p$ for $E=1$, $\sigma_1=10$, $\sigma_2=1$, $\theta_{1}=0.00001,0.2,0.4,0.5,0.6,0.8,1$ }
\label{fig:xp1}
\end{figure}

	If $E=1$ and, for example, $\sigma_1=10$, $\sigma_2=1$, and $\theta_1=0.0$, $0.2$, $0.4$, $0.5$, $0.6$, $0.8$, $1$, the behavior of $x_{0}$ for different values of $p$ can be observed in Fig~\ref{fig:xp1} and the values of $x_{0}$ for $p=1.1,1.3,1.6,2,2.7,4,10$ can be found in Table~1.

\begin{table}[h]
\label{table:root}
\begin{tabular}{|c|ccccccc|}
	\hline
$\theta_1$/$p$ &  $1.1$    &   $1.3$  & $1.6$   & $2$  &  $2.7$ & $4$ & $10$ \\
	\hline
0.0    & -0.00 & -0.00 & -0.00 & -0.00 & -0.00 & -0.00 & -0.00\\
0.2    & -0.25 & -0.24 & -0.21 & -0.18 & -0.14 & -0.10 & -0.05\\
0.4    & -0.67 & -0.61 & -0.52 & -0.43 & -0.34 & -0.25 & -0.12\\
0.5    & -0.99 & -0.87 & -0.72 & -0.60 & -0.47 & -0.35 & -0.17\\
0.6    & -1.43 & -1.19 & -0.98 & -0.82 & -0.65 & -0.48 & -0.24\\
0.8    & -2.33 & -2.03 & -1.74 & -1.50 & -1.24 & -0.96 & -0.52\\
1      & -3 & -3 & -3 & -3 & -3 & -3 & -3\\
	\hline
\end{tabular}
\caption{Values of $x_{0}$, when $E=1$, $\sigma_1=10$, and $\sigma_2=1$.}
\end{table}	

	In this case, observe that when $\theta_1=1$, $x_{0}=\frac{1-10}{3}=-3$.
	
\begin{figure}[h]
\centering
\includegraphics[width=0.5\columnwidth]{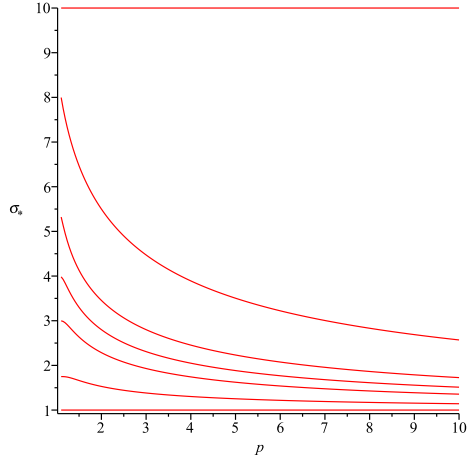}
\caption{Plot of $\sigma_{*}$ with respect to $p$ for $E=1$, $\sigma_1=10$, $\sigma_2=1$, $\theta_{1}=0.00001,0.2,0.4,0.5,0.6,0.8,1$ }
\label{fig:sp1}
\end{figure}

	Since $\sigma_*=\sigma_2(1-3x_{0})$, we have $$\frac{\partial\sigma_{*}}{\partial p}=-3\sigma_{2}\frac{\partial x_0}{\partial p}\leq0,$$ and it is equal to $0$ if $A=0$ which corresponds to the case when $\theta_{1}=1$.  So $\sigma_{*}$ is an strictly decreasing function of $p$, except when $\theta_{1}=1$ (all nonlinear material), in which case it is a constant function $\sigma_{*}=\sigma_1$ (=$10$ with the values given previously).  The behavior of $\sigma_{*}$ for different values of $p$ can be observed in Fig~\ref{fig:sp1} and the values of $\sigma_*$ for $p=1.1,1.3,1.6,2,2.7,4,10$ can be found in Table~2.  
\begin{table}[h]
\label{table:cond}
\begin{tabular}{|c|ccccccc|}
	\hline
$\theta_1$/$p$ &  $1.1$    &   $1.3$  & $1.6$   & $2$  &  $2.7$ & $4$ & $10$ \\
	\hline
0.0    & 1.00 & 1.00 & 1.00 & 1.00 & 1.00 & 1.00 & 1.00\\
0.2    & 1.75 & 1.72 & 1.63 & 1.53 & 1.42 & 1.30 & 1.14\\
0.4    & 2.99 & 2.84 & 2.55 & 2.29 & 2.01 & 1.74 & 1.36\\
0.5    & 3.98 & 3.62 & 3.17 & 2.80 & 2.42 & 2.05 & 1.51\\
0.6    & 5.29 & 4.57 & 3.94 & 3.45 & 2.94 & 2.45 & 1.72\\
0.8    & 7.99 & 7.09 & 6.22 & 5.50 & 4.72 & 3.89 & 2.57\\
1      & 10 & 10 & 10 & 10 & 10 & 10& 10\\
	\hline
\end{tabular}
\caption{Values of $\sigma_*$, when $E=1$, $\sigma_1=10$, and $\sigma_2=1$.}
\end{table}	

\begin{figure}[h]
\centering
\includegraphics[width=0.5\columnwidth]{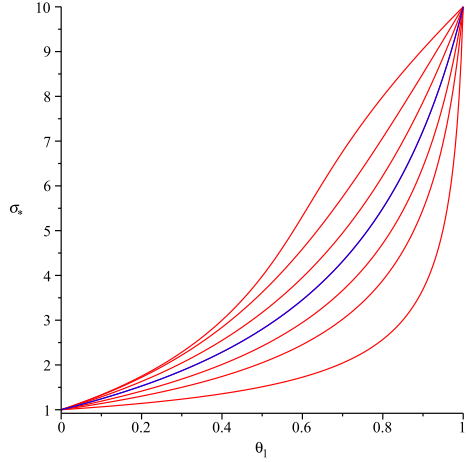}
\caption{Plot of $\sigma_{*}$ with respect to $\theta_{1}$ for $E=1$, $\sigma_1=10$, $\sigma_2=1$, $p=1.1,1.3,1.6,2,2.7,4,10$.  The blue plot corresponds to $p=2$ (Hashin-Shtrikman curve), above this curve are the plots for $p<2$ and below for $p>2$ in decreasing and increasing order respectively.}
\label{fig:st1}
\end{figure}

	Also, as a function of $\theta_{1}$, the effective conductivity $\sigma_*$ satisfies $$\frac{\partial\sigma_{*}}{\partial\theta_{1}}=-3\sigma_2\frac{x_{0}}{\theta_{1}^2}\left(\frac{\sigma_1(p-1)(1+Ax_{0})^{p-2}+2\sigma_2}{\sigma_1(p-1)A(1+Ax_0)^{p-2}+B\sigma_2}\right)>0.$$  Therefore $\sigma_*$ is an increasing function with respect to $\theta_1$.  The plot of $\sigma_*$ varying with respect to the volume fraction $\theta_{1}$ in the case when $E=1$, $\sigma_1=10$, and $\sigma_2=1$ can be observed in Fig~\ref{fig:st1}.  The blue plot corresponds to the case when $p=2$ (Hashin-Shtrikman lower bound for different values of $\theta_{1}$).  All the curves have fixed values $\sigma_*=\sigma_2$ for $\theta_{1}=0$ and $\sigma_*=\sigma_1$ for $\theta_{1}=1$.  In the case when $\sigma_1=10$, and $\sigma_2=1$, then $\sigma_*=1$ for $\theta_{1}=0$ and $\sigma_*=10$ for $\theta_{1}=1$ (see Fig~\ref{fig:st1}).

	We now investigate the cases when $0<E<1$ and $E>1$.  For these cases, we have that 
	\begin{equation}
	\label{xpE}
	\displaystyle
	\frac{\partial x_{0}}{\partial p}=\frac{-\sigma_{1}(E+Ax_{0})^{p-1}\ln(E+Ax_{0})}{\sigma_1(p-1)A(E+Ax_{0})^{p-2}+\sigma_2B}.
	\end{equation}	
Also, since $\sigma_*=\frac{\sigma_2}{E}(E-3x_{0})$, we have
	\begin{equation}
	\label{spE}
	\displaystyle
	\frac{\partial\sigma_{*}}{\partial p}=-3\frac{\sigma_{2}}{E}\frac{\partial x_0}{\partial p}.
	\end{equation}
With respect to $\theta_1$, we obtain 
	\begin{equation}
	\label{xtE}
\displaystyle	\frac{\partial x_{0}}{\partial\theta_1}=\frac{x_{0}}{\theta_{1}^2}\left[\frac{\sigma_{1}(p-1)(E+Ax_{0})^{p-2}+2\sigma_2}{\sigma_1(p-1)A(E+Ax_{0})^{p-2}+\sigma_2B}\right],
\end{equation} 
and the change of the effective conductivity $\sigma_*$ with respect to the volume fraction $\theta_{1}$ is given by	
\begin{equation}
\label{stE}
\displaystyle	\frac{\partial\sigma_*}{\partial\theta_1}=\frac{-3\sigma_2x_{0}}{E\theta_{1}^2}\left[\frac{\sigma_{1}(p-1)(E+Ax_{0})^{p-2}+2\sigma_2}{\sigma_1(p-1)A(E+Ax_{0})^{p-2}+\sigma_2B}\right].
\end{equation} 

\begin{figure}[h]
\centering
\includegraphics[width=0.5\columnwidth]{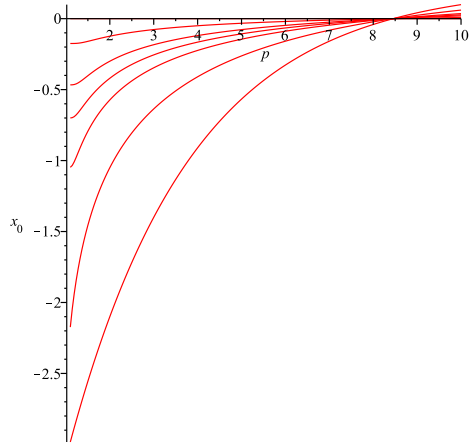}
\caption{Plot of $x_{0}$ with respect to $p$ for $E=0.7$, $\sigma_1=10$, $\sigma_2=1$, $\theta_{1}=0.00001,0.2,0.4,0.5,0.6,0.8,1$ }
\label{fig:xp07}
\end{figure}

	We start with $0<E<1$.  In this case, since $0<E+Ax_0<1$, we have from (\ref{xpE}) that $\displaystyle \frac{\partial x_{0}}{\partial p}>0$, and we conclude $x_{0}$ is an strictly increasing function of $p$.  If $E=0.7$, $\sigma_1=10$, $\sigma_2=1$, and $\theta_1=0.0,0.2,0.4,0.5,0.6,0.8,1$, the behavior of $x_{0}$ for different values of $p$ can be observed in Fig~\ref{fig:xp07}; and the values of $x_{0}$ for $p=1.1,1.3,1.6,2,2.7,4,10$ can be found in Table~3.
\begin{table}
\label{table:root07}
\begin{tabular}{|c|ccccccc|}
	\hline
$\theta_1$/$p$ &  $1.1$    &   $1.3$  & $1.6$   & $2$  &  $2.7$ & $4$ & $10$ \\
	\hline
0.0    & 0.00 & 0.00 & 0.00 & 0.00 & 0.00 & 0.00 & 0.00\\
0.2    & -0.17 & -0.17 & -0.15 & -0.12 & -0.09 & -0.05 &  0.01\\
0.4    & -0.47 & -0.45 & -0.38 & -0.30 & -0.21 & -0.12 &  0.02\\
0.5    & -0.70 & -0.65 & -0.54 & -0.42 & -0.29 & -0.17 &  0.03\\
0.6    & -1.04 & -0.92 & -0.74 & -0.57 & -0.40 & -0.22 &  0.04\\
0.8    & -2.17 & -1.73 & -1.35 & -1.05 & -0.73 & -0.41 &  0.06\\
1      & -2.99 & -2.75 & -2.45 & -2.10 & -1.58 & -0.91 &  0.10\\
	\hline
\end{tabular}
\caption{Values of $x_{0}$, when $E=0.7$, $\sigma_1=10$, and $\sigma_2=1$.}
\end{table} 

\begin{figure}[h]
\centering
\includegraphics[width=0.5\columnwidth]{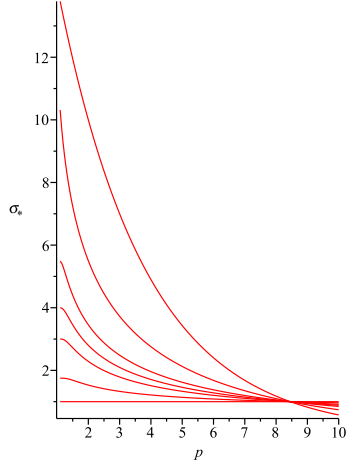}
\caption{Plot of $\sigma_{*}$ with respect to $p$ for $E=0.7$, $\sigma_1=10$, $\sigma_2=1$, $\theta_{1}=0.00001,0.2,0.4,0.5,0.6,0.8,1$ }
\label{fig:sp07}
\end{figure}

  By (\ref{spE}), we can also conclude that $\sigma_{*}$ is an strictly decreasing function of $p$.  The behavior of $\sigma_{*}$ for different values of $p$, in the case when $E=0.7$, $\sigma_1=10$, and $\sigma_2=1$, can be observed in Fig~\ref{fig:sp07} and the values of $\sigma_*$ for $p=1.1,1.3,1.6,2,2.7,4,10$ can be found in Table~4.
\begin{table}[h]
\label{table:cond07}
\begin{tabular}{|c|ccccccc|}
	\hline
$\theta_1$/$p$ &  $1.1$    &   $1.3$  & $1.6$   & $2$  &  $2.7$ & $4$ & $10$ \\
	\hline
0.0    & 1.00 & 1.00 & 1.00 & 1.00 & 1.00 & 1.00 & 1.00\\
0.2    & 1.75 & 1.74 & 1.66 & 1.53 & 1.38 & 1.22 & 0.96\\
0.4    & 3.00 & 2.93 & 2.62 & 2.29 & 1.90 & 1.52 & 0.92\\
0.5    & 4.00 & 3.80 & 3.30 & 2.80 & 2.26 & 1.72 & 0.89\\
0.6    & 5.47 & 4.96 & 4.15 & 3.46 & 2.70 & 1.96 & 0.85\\
0.8    & 10.3 & 8.42 & 6.79 & 5.51 & 4.13 & 2.76 & 0.74\\
1      & 13.8 & 12.8 & 11.5 & 10 & 7.78 & 4.90& 0.58\\
	\hline
\end{tabular}
\caption{Values of $\sigma_*$, when $E=0.7$, $\sigma_1=10$, and $\sigma_2=1$.}
\end{table}
	  
\begin{figure}[h]
\centering
\includegraphics[width=0.5\columnwidth]{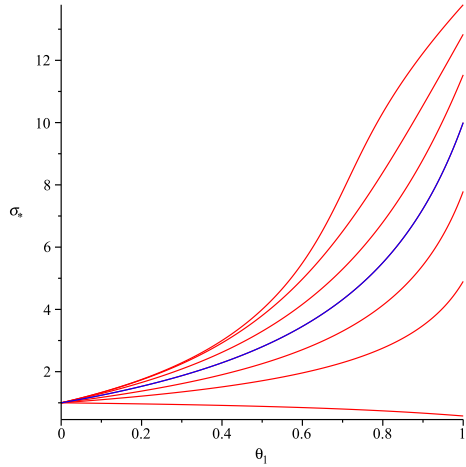}
\caption{Plot of $\sigma_{*}$ with respect to $\theta_{1}$ for $E=0.7$, $\sigma_1=10$, $\sigma_2=1$, $p=1.1,1.3,1.6,2,2.7,4,10$.  The blue plot corresponds to $p=2$.}
\label{fig:st07}
\end{figure}

	The rate of change of the effective conductivity $\sigma_*$ with respect to the volume fraction $\theta_{1}$ (\ref{stE}) is determined by the sign of $x_0$.  If $x_{0}<0$, then $\sigma_*$ is an strictly increasing function of $\theta_1$, and if $x_{0}>0$ it is decreasing.  We obtain have that $\sigma_{*}$ is a constant function of $\theta_{1}$ if $\displaystyle \frac{\sigma_1}{\sigma_{2}}=\frac{1}{E^{p-2}}$.  A plot of $\sigma_*$ as a function of $\theta_1$ can be observed in Fig~\ref{fig:st07} for the values $E=0.7$, $\sigma_1=10$, and $\sigma_2=1$.  The blue plot corresponds to the case when $p=2$ (Hashin-Shtrikman lower bound for different values of $\theta_{1}$). The curves have values $\sigma_*=\sigma_2=1$ for $\theta_{1}=0$ and $\sigma_*=\sigma_1\left|E\right|^{p-2}=10(0.7)^{p-2}$ for $\theta_{1}=1$.  Observe for example, for $p=1.1$ and $\theta_{1}=1$, $\sigma_*=10(0.7)^{-0.9}\approx13.8$ and for $p=10$, $\sigma_*=10(0.7)^{8}\approx0.58$.  Observe in Fig~\ref{fig:st07} that for $p=10$, the effective conductivity $\sigma_*$ is decreasing with respect to $\theta_1$, this is obtained from the fact that $x_{0}>0$ for $p=10$, $E=0.7$, $\sigma_1=10$, and $\sigma_2=1$. 

	When $E>1$, we have $\displaystyle \frac{\partial x_{0}}{\partial p}\geq0$ iff $\displaystyle \sigma_{1}\geq\sigma_{2}\left(\frac{3E-2-\theta_1}{1-\theta_1}\right),$ otherwise $\displaystyle \frac{\partial x_{0}}{\partial p}<0$.  For example, if $E=2$, $\sigma_1=10$, $\sigma_2=1$, and $\theta_1=0.0$, $0.2$, $0.4$, $0.5$, $0.6$, $0.8$, $1$, the values of $x_{0}$ for $p=1.1$, $1.3$, $1.6$, $2$, $2.7$, $4$, $10$ can be found in Table~5.  Observe that, for instance, for $\theta_{1}=0.4$, $x_0$ is increasing as $p$ increases, but for $\theta_{1}=0.8$, $x_{0}$ is decreasing as $p$ increases.
  
\begin{table}[h]
\label{table:root2}
\begin{tabular}{|c|ccccccc|}
	\hline
$\theta_1$/$p$ &  $1.1$    &   $1.3$  & $1.6$   & $2$  &  $2.7$ & $4$ & $10$ \\
	\hline
0.0    & -0.00 & -0.00 & -0.00 & -0.00 & -0.00 & -0.00 & -0.00\\
0.2    & -0.49 & -0.43 & -0.39 & -0.35 & -0.32 & -0.30 & -0.27\\
0.4    & -1.14 & -1.01 & -0.92 & -0.86 & -0.80 & -0.76 & -0.70\\
0.5    & -1.48 & -1.35 & -1.26 & -1.20 & -1.14 & -1.09 & -1.04\\
0.6    & -1.79 & -1.73 & -1.67 & -1.64 & -1.59 & -1.56 & -1.52\\
0.8    & -2.38 & -2.56 & -2.78 & -3. & -3.25 & -3.5 & -3.81\\
1      & -2.90 & -3.43 & -4.40 & -6. & -10.1 & -26. &  -1710\\
	\hline
\end{tabular}
\caption{Values of $x_{0}$, when $E=2$, $\sigma_1=10$, and $\sigma_2=1$.}
\end{table}

	The values of $\sigma_*$ for $p=1.1,1.3,1.6,2,2.7,4,10$ can be found in Table~6.  By (\ref{spE}), we have $\sigma_*$ it is increasing iff $\displaystyle \sigma_{1}<\sigma_{2}\left(\frac{3E-2-\theta_1}{1-\theta_1}\right),$ otherwise $\displaystyle \frac{\partial\sigma_{*}}{\partial p}\geq0$.   For example, for $\theta_{1}=0.4$, the value of $\sigma_{*}$ decreases as $p$ increases but for $\theta_{1}=0.8$, $\sigma_*$ increases as $p$ increases.   
\begin{table}[h]
\label{table:cond2}
\begin{tabular}{|c|ccccccc|}
	\hline
$\theta_1$/$p$ &  $1.1$    &   $1.3$  & $1.6$   & $2$  &  $2.7$ & $4$ & $10$ \\
	\hline
0.0    & 1.00 & 1.00 & 1.00 & 1.00 & 1.00 & 1.00 & 1.00\\
0.2    & 1.73 & 1.65 & 1.58 & 1.53 & 1.48 & 1.44 & 1.40\\
0.4    & 2.71 & 2.52 & 2.38 & 2.28 & 2.20 & 2.13 & 2.05\\
0.5    & 3.22 & 3.02 & 2.89 & 2.80 & 2.71 & 2.64 & 2.56\\
0.6    & 3.68 & 3.60 & 3.50 & 3.46 & 3.38 & 3.34 & 3.28\\
0.8    & 4.57 & 4.84 & 5.15 & 5.50 & 5.90 & 6.25 & 6.71\\
1      & 5.35 & 6.15 & 7.60 & 10 & 16.2 & 40.0 & 2560\\
	\hline
\end{tabular}
\caption{Values of $\sigma_*$, when $E=2$, $\sigma_1=10$, and $\sigma_2=1$.}
\end{table}

\begin{figure}[h]
\centering
\includegraphics[width=0.5\columnwidth]{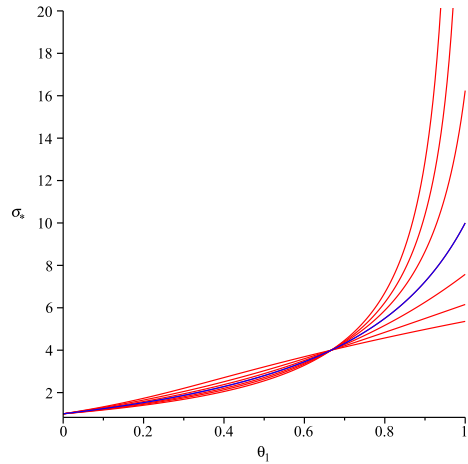}
\caption{Plot of $\sigma_{*}$ with respect to $\theta_{1}$ for $E=2$, $\sigma_1=10$, $\sigma_2=1$, $p=1.1,1.3,1.6,2,2.7,4,10$.  The blue plot corresponds to $p=2$.}
\label{fig:st2}
\end{figure}

	Observe that $f(0)=\sigma_1E^{p-1}-\sigma_2E>0$ if $E>1$, therefore $x_{0}<0$ and, by (\ref{stE}), the effective conductivity $\sigma_*$ is an strictly increasing function of $\theta_{1}$.  The effective conductivity $\sigma_*$ as a function of the volume fraction $\theta_{1}$ for the case when $E=2$, $\sigma_1=10$, and $\sigma_2=1$ can be observed in Fig~\ref{fig:st2}.  The blue plot corresponds to the case when $p=2$ (Hashin-Shtrikman lower bound for different values of $\theta_{1}$).  

\section{Final Remarks}
\label{Fin}
We considered the problem of constructing neutral inclusions from nonlinear materials.  In particular, we studied the case of a coated sphere in which the core was nonlinear and the coating was a linear material.  We showed that the coated sphere is equivalent to a sphere composed only of linear material of conductivity $\sigma_{*}$ (\ref{7}).  Thus the coated sphere is neutral in an environment formed by a linear material of conductivity $\sigma_{*}$.  One could continue to add coated spheres of various sizes but with fixed volume fraction $\theta_{1}$, without disturbing the prescribed uniform applied electric field surrounding the inclusions, fill the entire space with an assemblage of these coated spheres and this configuration of nonlinear materials dissipates energy the same as a linear material with thermal conductivity $\sigma_{*}$.  We then studied the two-dimensional case of assemblages of coated disks.

Ongoing work includes the generalization of this work to assemblages of coated ellipsoids, and to investigate if the coated spheres (disks) assemblage constructed in this paper could be an optimal microstructure, in the sense that these microstructures attain bounds for the effective conductivity of a composited made of a linear matrix filled with a p-harmonic material.

\section{ACKNOWLEDGMENT}
The authors would like to thank Robert P. Lipton for fruitful discussions and helpful suggestions.
	
\bibliographystyle{alpha}	
\bibliography{thesis}
\end{document}